\documentclass[aps,prb,preprint,showkeys]{revtex4-1}
\usepackage{geometry}

\usepackage{graphicx}
\usepackage{tikz}
\usetikzlibrary{decorations.pathmorphing}
\usepackage[T1]{fontenc}
\usepackage{booktabs} 
\usepackage{mathrsfs}
\usepackage{miller}
\usepackage{amsmath} 
\usepackage{amssymb}
\usepackage{amsfonts}
\usepackage{xspace}
\usepackage{siunitx}
\usepackage{hyperref} 
\usepackage[caption=false]{subfig}

\DeclareSIUnit\electrons{e\textsuperscript{-}}
\DeclareSIUnit\neutrons{neutrons}
\DeclareSIUnit\ppm{ppm}
\DeclareSIUnit\ppb{ppb}
\DeclareSIUnit\lines{l}
\DeclareSIUnit{\calorie}{cal}

\newcommand{\symspc}[1]{\ensuremath{\mathrm{#1}}}

\newcommand{\Vneutral}{\ensuremath{\mathrm{V}^{0}}\xspace}

\newcommand{\Cthir}{\ensuremath{^{13}\mathrm{C}}\xspace}

\newcommand{\Nfif}{\ensuremath{^{15}\mathrm{N}}\xspace}
\newcommand{\Nfour}{\ensuremath{^{14}\mathrm{N}}\xspace}

\newcommand{\NfifNVNminus}{\ensuremath{\mathrm{^{15}N_{2}V^{-}}}}
\newcommand{\NfourNVNminus}{\ensuremath{\mathrm{^{14}N_{2}V^{-}}}}

\newcommand{\PtwoConsNeutral}{\ensuremath{\mathrm{N}_{3}\mathrm{V}^{0}}\xspace}

\newcommand{\BConsNeutral}{\ensuremath{\mathrm{N}_{4}\mathrm{V}^{0}}\xspace}

\newcommand{\NVNnb}{\ensuremath{\mathrm{N_{2}V}}\xspace} 
\newcommand{\NVnb}{\ensuremath{\mathrm{NV}}\xspace}

\newcommand{\NVminus}{\ensuremath{\NVnb^{-}}\xspace}
\newcommand{\NVneutral}{\ensuremath{\NVnb^{0}}\xspace}

\newcommand{\NVNminus}{\ensuremath{\mathrm{N_{2}V}^{-}}\xspace}
\newcommand{\NVNneutral}{\ensuremath{\mathrm{N_{2}V}^{0}}\xspace}
\newcommand{\NVN}{\NVNnb\xspace} 

\newcommand{\Ns}{\ensuremath{\mathrm{N_{s}}}\xspace}

\newcommand{\Nsneutral}{\ensuremath{\mathrm{N_{s}}^{0}}\xspace}
\newcommand{\NsPlus}{\ensuremath{\mathrm{N_{s}}^{+}}\xspace}
\newcommand{\NfourNSub}{\ensuremath{^{14}\mathrm{N_{s}}^{0}}\xspace}

\newcommand{\NfifNSub}{\ensuremath{^{15}\mathrm{N}_{\mathrm{s}}^{0}}\xspace}
\newcommand{\NfifNSubPlus}{\ensuremath{^{15}\mathrm{N}_{\mathrm{s}}^{+}}\xspace}

\newcommand{\RhombicI}{\ensuremath{\mathscr{C}_{\mathrm{2v}}}\xspace}

\makeatletter
\def\@author#1{\g@addto@macro\elsauthors{\normalsize%
    \def\baselinestretch{1}%
    \upshape\authorsep#1\unskip\textsuperscript{%
      \ifx\@fnmark\@empty\else\unskip\sep\@fnmark\let\sep=,\fi
      \ifx\@corref\@empty\else\unskip\sep\@corref\let\sep=,\fi
      }%
    \def\authorsep{\unskip,\space}%
    \global\let\@fnmark\@empty
    \global\let\@corref\@empty  
    \global\let\sep\@empty}%
    \@eadauthor={#1}
}
\makeatother

\begin{document}

\title{Electron paramagnetic resonance of the \texorpdfstring{$\mathbf{N_{2}V^{-}}$}{(NVN)-} defect in \texorpdfstring{$\mathbf{^{15}N}$}{15N}-doped synthetic diamond}

\author{B.\ L.\ Green}
\altaffiliation{Corresponding Author}
\email{b.green@warwick.ac.uk}

\author{M.\ W.\ Dale}

\author{M.\ E.\ Newton}
\affiliation{Department of Physics, University of Warwick, Coventry, CV4 7AL,  United Kingdom}

\author{D.\ Fisher}
\affiliation{De Beers Technologies, Maidenhead, Berkshire, SL6 6JW, United Kingdom}

\begin{abstract}
Nitrogen is the dominant impurity in the majority of natural and synthetic diamonds, and the family of nitrogen vacancy-type ($\mathrm{N_{n}V}$) defects are crucial in our understanding of defect dynamics in these diamonds. A significant gap is the lack of positive identification of \NVNminus, the dominant charge state of $\mathrm{N_{2}V}$ in diamond that contains a significant concentration of electron donors. In this paper we employ isotopically-enriched diamond to identify the EPR spectrum associated with \NfifNVNminus and use the derived spin Hamiltonian parameters to identify \NfourNVNminus{} in a natural isotopic abundance sample. The electronic wavefunction of the \NVNminus ground state and previous lack of identification is discussed. The \NVNminus EPR spectrum intensity is shown to correlate with H2 optical absorption over an order of magnitude in concentration.
\end{abstract}

\maketitle

\section{Introduction}
\label{sec:introduction}

Nitrogen is the most common impurity found in most natural and synthetic diamonds (with the possible exception of hydrogen), and nitrogen-containing complexes have attracted significant research interest during the past 50~years. A significant proportion of recent research into diamond has concentrated primarily on the quantum optoelectronic properties of the negatively-charged nitrogen-vacancy (\NVminus, a substitutional nitrogen atom nearest-neighbor to a vacant lattice site) defect and its potential use in quantum metrology and sensing.\cite{Acosta2013,Schirhagl2014,Doherty2013,Rondin2014} \NVneutral has been well characterized by optical\cite{Davies1976a,Collins1987a} and electron paramagnetic resonance (EPR)\cite{Felton2008} techniques: the relative concentrations of \NVneutral and \NVminus are determined by the availability of a suitable electron donor (e.g.\ single substitutional nitrogen \Ns).

\begin{figure*}
	\subfloat[]{\includegraphics[width=0.4\textwidth]{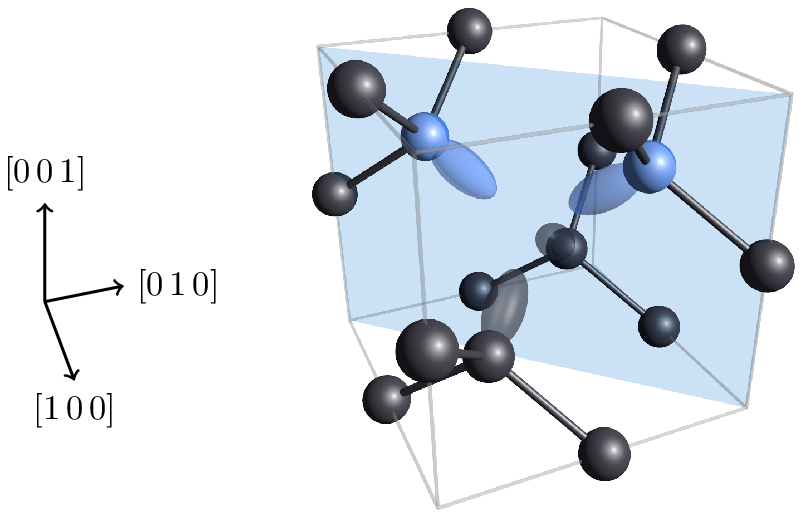}\label{subfig:NVNdrawing}}
	\qquad \qquad
	\subfloat[]{\includegraphics[width=0.4\textwidth]{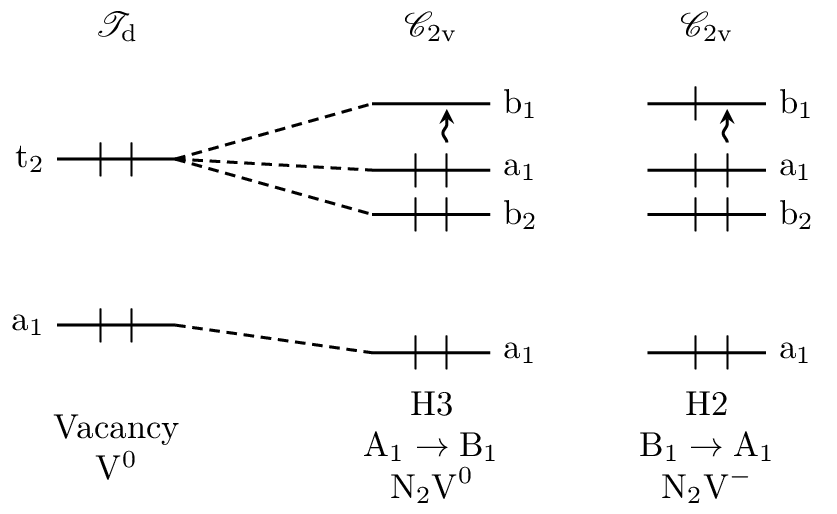}\label{subfig:N2Velectronic}}
	\caption{(a): the structure of the \NVN{} defect in diamond. Blue (light) atoms are nitrogen, and nominal $sp$-type orbitals are drawn for illustration. The defect has \RhombicI symmetry; the \hkl(1-10) mirror plane is shown. (b): the effective one-electron picture produced by starting with the structure of the vacancy, lowering the symmetry and adding two (three) electrons for H3 (H2) using the procedure described by Lowther.\cite{Lowther1984} Electronic occupancy of orbitals indicated by vertical bars, with arrows indicating electronic absorption transitions between ground and excited states of symmetry \symspc{A_{1}} or \symspc{B_{1}}. Adapted from Lawson.\cite{Lawson1992}}
	\label{subfig:NVN_structure}
\end{figure*}

$\mathrm{NV^{0/-}}$ belong to the $\mathrm{N_{n}V}$ family of defects, where $\mathrm{n} = \SIrange{1}{4}{}$: the defects \NVNneutral, \PtwoConsNeutral and  \BConsNeutral are identified with the H3 (zero phonon line (ZPL) \SI{2.465}{\electronvolt}),\cite{Davies1976b,Davies1976} N3 (\SI{2.985}{\electronvolt})\cite{Thomaz1978} and B-center (infrared absorption band) optical bands that have been extensively studied.\cite{Chrenko1977,Collins1980,Evans1982,Goss2001} A notable void in the understanding of $\mathrm{N_{n}V}$-type complexes is the lack of identification of \NVNminus (see figure~\ref{subfig:NVNdrawing}): a number of candidate optical and paramagnetic signatures have been postulated,\cite{Lawson1992,Nisida1992} the most promising of which is the optical H2 band, however no conclusive identification has been made of the defect by EPR.

The H2 optical band was first reported in 1956 as a broad absorption feature observed after irradiation and annealing of natural diamond.\cite{Clark1956a} At \SI{80}{\kelvin}, H2 is observed in both luminescence and absorption with a ZPL at \SI{1.257}{\electronvolt} and an accompanying vibronic band. Uniaxial stress measurements of the H2 ZPL assigned a symmetry of \RhombicI,\cite{Lawson1992} while photochromism charge balance studies suggest that the defect is the negative charge state of the H3 optical defect.\cite{Mita1990} 

The assignment of H2 and H3 to different charge states of the same defect presents a potential problem: the ground state of \NVNminus should be EPR-active (with spin $S = \frac{1}{2}$), but has not been observed.\cite{Lawson1992} One candidate EPR spectrum has been tentatively attributed to \NVNminus, but the given spin Hamiltonian parameters are a poor fit to the limited experimental data.\cite{Nisida1992}

The electronic structure of \NVNneutral/\NVNminus (see figure~\ref{subfig:N2Velectronic}) precludes any simple spin-polarization mechanisms such as those seen in \NVminus and $\mathrm{SiV^{0}}$;\cite{Doherty2013} however the known one-electron ionization behaviour\cite{Mita1990,Mita1993} from \NVNminus (paramagnetic) to \NVNneutral (diamagnetic) makes the \NVN system a candidate for long-lived nuclear spin memory protocols such as those recently exploited in silicon.\cite{Saeedi2013}

In this paper, an EPR spectrum is identified with \NfifNVNminus{} in a treated \Nfif-enriched diamond; \Nfif and \Cthir spin Hamiltonian parameters are extracted from the EPR spectrum and used to aid identification of \NfourNVNminus{} in a sample with C \& N isotopes in natural abundance. The spin Hamiltonian parameters are discussed and are employed to illustrate the difficulty in identifying the spectrum in natural abundance samples.

\section{Synthesis of isotopically enriched diamond}
\label{sec:synthesis}
Isotopic enrichment is a well-established technique in the study of defects in diamond: nitrogen enrichment has been employed for over 30 years;\cite{Collins1982c} carbon enrichment for at least 20.\cite{Anthony1990} Recent reports have concentrated on carbon isotopic enrichment in chemical vapor deposition (CVD)-grown diamond; however CVD synthesis becomes problematic with high levels of gas-phase $\mathrm{N}_{2}$ and hence high pressure high temperature (HPHT) is still preferred when doping with $\gtrapprox \SI{10}{\ppm}$ nitrogen in the solid phase.

Atmospheric nitrogen adsorbed (absorbed) onto (into) the growth source and capsule materials is readily incorporated into the synthesized diamond ($[\Nsneutral] \gtrapprox \SI{100}{\ppm}$ are typical) if preventative measures are not employed. Such measures include chemical nitrogen traps (so-called ``getters'') added to the growth materials, or outgassing all growth materials pre-synthesis and sealing the growth capsule. The sample used in this experiment was grown by evacuating and outgassing a sealed HPHT synthesis capsule and subsequently backfilling with \Nfif-enriched $\mathrm{N_{2}}$ gas:\cite{Stromann2006} $\gtrapprox \SI{95}{\percent}$ of incorporated nitrogen was \Nfif{} --- nominally identical to the source gas.

\section{Experimental detail}
\label{sec:experiment}

\subsection{Sample}
\label{subsec:sample}
Both \Nsneutral and A-centers ($(\Ns-\Ns)^{0}$) are effective traps for mobile vacancies. If \Nsneutral is the most abundant impurity post-irradiation (to produce vacancies and interstitials; we will not concern ourselves here with the interaction of self-interstitials with nitrogen defects nor the charge state of the vacancy, as neither has a bearing on the following discussion), then upon annealing to temperatures where the vacancy is mobile ($\gtrapprox \SI{600}{\celsius}$) the dominant aggregation mechanisms will be $\Nsneutral + \mathrm{V} \rightarrow \NVneutral$ and $\Nsneutral + \NVneutral \rightarrow \NsPlus + \NVminus$; similarly if A-centers are abundant then the dominant process becomes $(\Ns-\Ns)^{0} + \mathrm{V} \rightarrow \NVNneutral$. If both A-centers and \Nsneutral are present they will both trap mobile vacancies and \Nsneutral will donate charge to $\mathrm{N_{2}V}$, undergoing the additional process $\Nsneutral + \NVNneutral \rightarrow \NsPlus + \NVNminus$. NV is stable to approximately \SI{1500}{\celsius}; at this temperature the reaction $\Nsneutral + \NVneutral \rightarrow \NVNneutral$ and of course $\Nsneutral + \NVNneutral \rightarrow \NsPlus + \NVNminus$ start to become significant. $\mathrm{N_{2}V}$ is itself only stable to approximately \SI{1600}{\celsius} and anneals out with the production of A-centers ($\NVN \rightarrow (\Ns-\Ns) + \mathrm{V}$), where the vacancy is recycled to promote further nitrogen aggregation / migration.\cite{Collins1980} Thus when irradiated type~Ib (\Nsneutral is the dominant impurity in starting material) diamond is annealed, only a very narrow window exists where $\mathrm{N_{2}V}$ is produced over NV --- it is challenging to produce large concentrations of $\mathrm{N_{2}V}$ by this route. However, if the type~Ib diamond is subjected to HPHT treatment to convert a substantial fraction of \Nsneutral (but not all) to A-centers before irradiation and subsequent annealing at \SI{800}{\celsius}, then significant concentrations of \NVNminus (and \NVNneutral) can be produced.

\begin{figure*}[hbt]
	\subfloat[]{\label{subfig:SYN339UVVis}\includegraphics[height=14em]{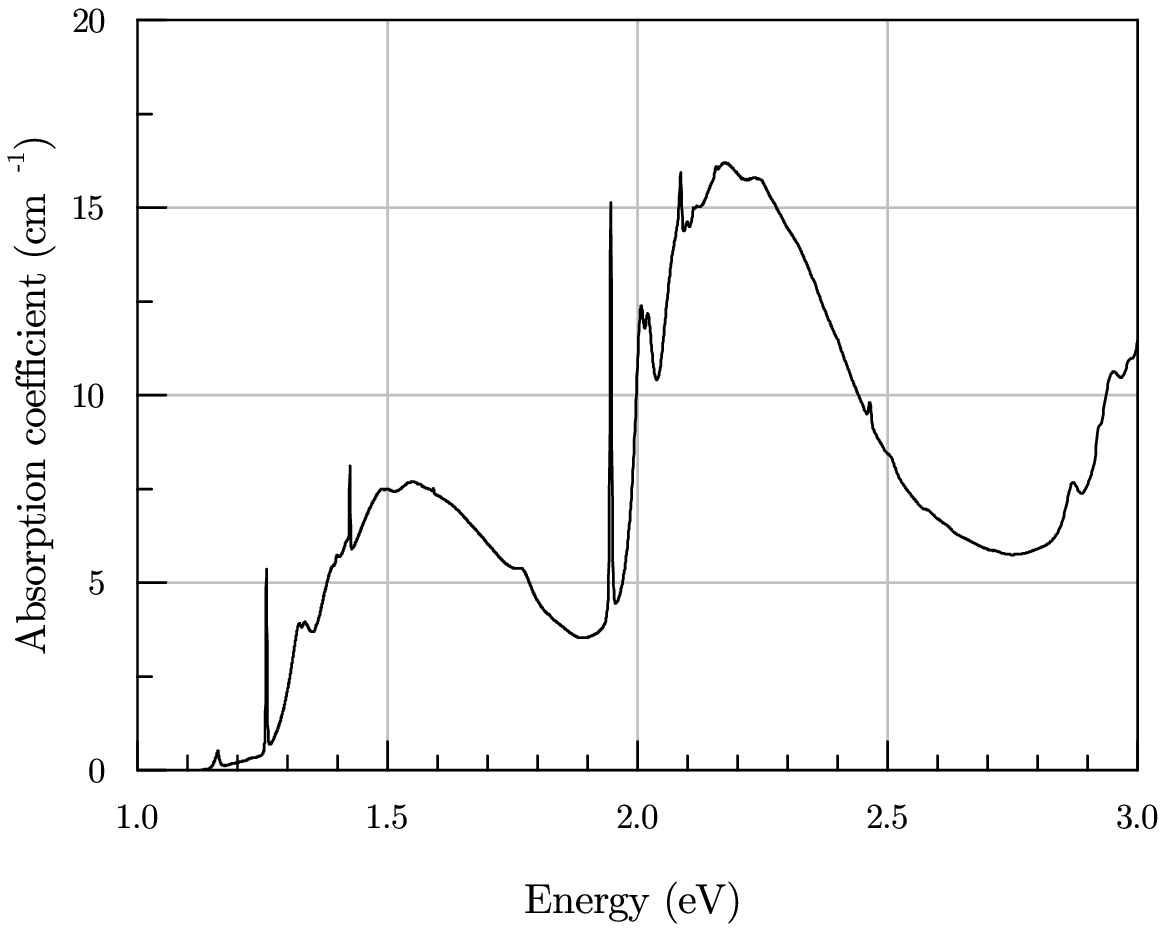}}
	\qquad
	\subfloat[]{\label{subfig:h2vsN2V}\includegraphics[height=14em]{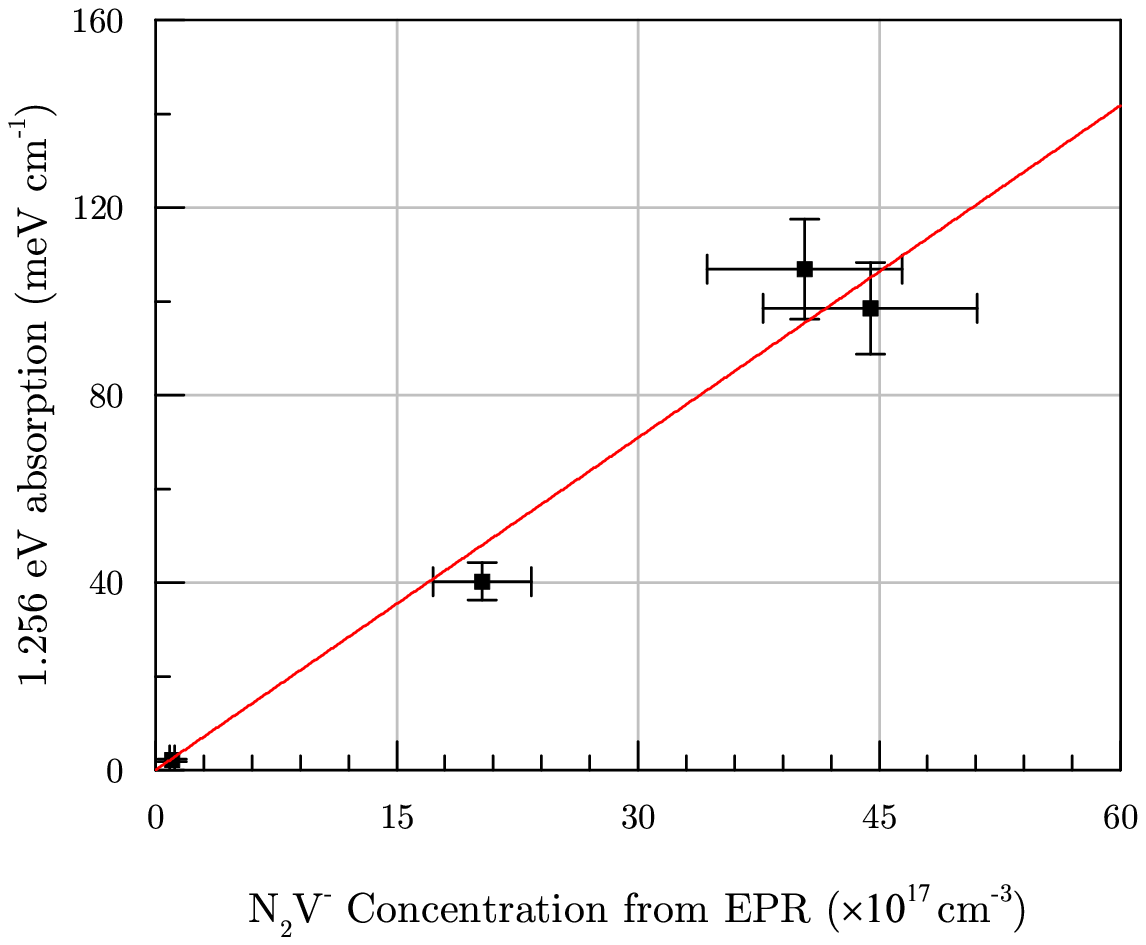}}
	\caption{(a): NIR-visible absorption spectrum of the sample following high temperature high pressure annealing, electron irradiation and further annealing. Strong absorption bands include H2 (\SI{1.257}{\electronvolt}) and \NVminus (\SI{1.945}{\electronvolt}). (b): correlation between the integrated absorption of the H2 optical band and the \NfifNVNminus{} EPR intensity in neutron-irradiated and annealed samples.}
\end{figure*}

The sample used in this study was annealed using the above protocol: it first underwent high temperature ($\gtrapprox \SI{1900}{\celsius}$) and pressure (of order \SI{5}{\giga\pascal}) treatment for \SI{1}{\hour} to produce nearest-neighbor nitrogen aggregates. The sample was then electron irradiated with \SI{4.5}{\mega\electronvolt} electrons to create vacancies. This processing produced approximately $[\Vneutral]=\SI{5+-1}{\ppm}$, and $[(^{15}\mathrm{N_{s}}-^{15}\mathrm{N_{s}})^{0}]=\SI{45+-3}{\ppm}$, $[\NfifNSub]=\SI{10+-1}{\ppm}$ and $[\NfifNSubPlus]=\SI{6+-2}{\ppm}$, accounting for pre-treatment concentration of $[\NfifNSub]=\SI{105+-5}{\ppm}$. Finally, the sample was further annealed at \SI{800}{\celsius} for \SI{14}{\hour} to produce nitrogen vacancy aggregates. Final paramagnetic defect concentrations were approximately $[\Nsneutral] = \SI{5.0+-0.4}{\ppm}$, $[\NVminus] = \SI{1.6+-0.3}{\ppm}$ and $[\NVNminus] = \SI{1.8+-0.2}{\ppm}$: the sample was notably photochromic and hence concentrations were subject to recent sample history. An NIR-visible absorption spectrum of the sample is given in figure~\ref{subfig:SYN339UVVis}.

\section{Results}
\label{sec:results}
A \RhombicI defect such as \NVN{} may be aligned along one of six equivalent orientations, each defined by its principal \hkl{1-10} plane (see Supplemental Material for illustrations of different orientations; all following directions are for the orientation given in figure~\ref{subfig:NVNdrawing}). We expect the $g$-tensor to be anisotropic with principal directions along \hkl[1-10], \hkl[110] and \hkl[001]. When observed in EPR with the Zeeman field $B$ along a \hkl<001> direction, the $g$-anisotropy splits the orientations into two distinct groups: two orientations with their defining \hkl<1-10> axis perpendicular to $B$; and four orientations with their axis at \SI{45}{\degree}. \NVNminus contains two equivalent nitrogen nuclei -- in this case both nuclei are \Nfif and therefore possess $I=\frac{1}{2}$. A reasonable assumption is that the \Nfif hyperfine interaction is approximately along the \hkl<111> direction connecting each nitrogen to the vacancy, and hence all nitrogen hyperfine interactions are equivalent when $B\|\hkl<001>$. The expected structure for an EPR spectrum recorded with $B\|\hkl<001>$ is therefore two sets of lines (with intensity 2:1 according to the 4:2 orientation ratio) each split by two equivalent $I=\frac{1}{2}$ nuclei.

\begin{figure}[tbh]
	\centering
	\includegraphics[width=0.9\columnwidth]{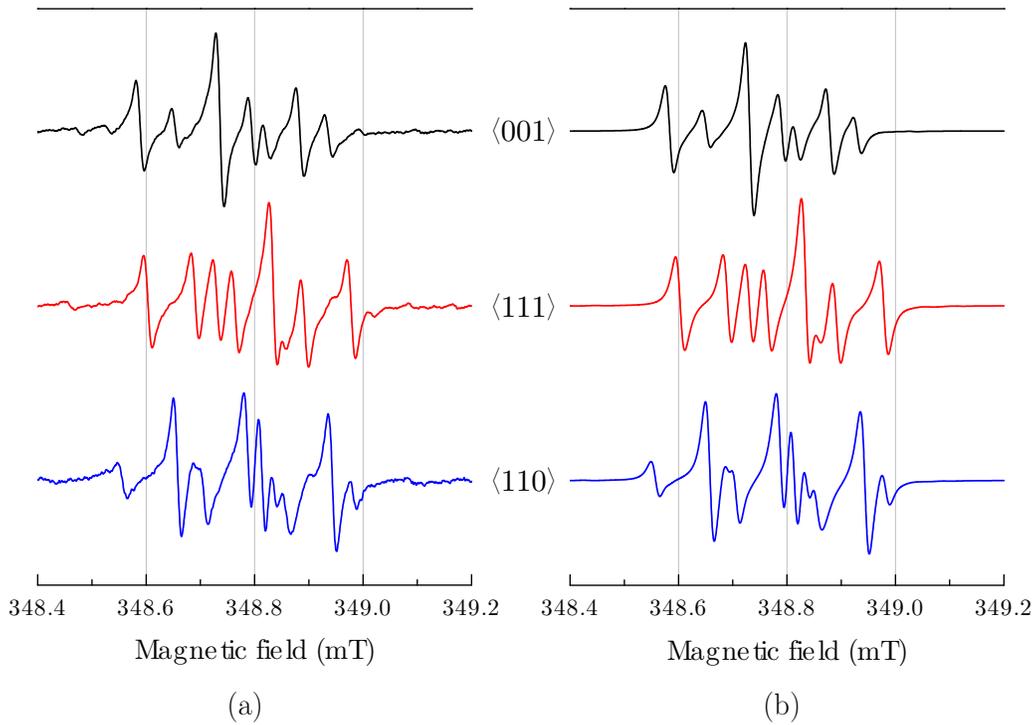}
	\caption{A comparison of experimental (a) and calculated spectra (b) at a microwave frequency of approximately \SI{9.755}{\giga\hertz}. Calculated spectra generated with EasySpin\cite{Stoll2006} using the spin Hamiltonian parameters determined by fitting the line positions (see table~\ref{tab:spin_hamiltonian_parameters}); EPR linewidth was adjusted to fit the experimental data. The orientation of the magnetic field for each spectrum is given in the center of the figure. The calculated spectra include \NfifNVNminus{} and \NfourNSub (visible at \SI{348.82}{\milli\tesla}), and have been referenced to \Nsneutral assuming that this defect possesses an isotropic $g$-value: \num{2.0024}.}
	\label{fig:EPRSpectrumComparison}
\end{figure}

The experimental $B\|\hkl<001>$ spectrum given in figure~\ref{fig:EPRSpectrumComparison} displays the expected structure: two sets of lines with relative intensities 1:2:1. Similar arguments apply for both $B\|\hkl<111>$ and $B\|\hkl<110>$ spectra. These spectra unambiguously identify an $S=\frac{1}{2}$ defect containing two equivalent $I=\frac{1}{2}$ nuclei at near \SI{100}{\percent} abundance, indicating \Nfif.

NIR-visible optical absorption measurements were performed at room temperature on several diamond samples treated with neutron irradiation and subsequent annealing. In all samples the intensity of the \NVNminus EPR spectrum was found to correlate with the integrated absorption of the H2 ZPL --- see figure~\ref{subfig:h2vsN2V}. This result gives additional evidence to the assignment of the H2 optical band to the \NVNminus defect.\cite{Mita1990,Lawson1992}

\begin{table}[tbh]
	\centering
	\footnotesize
	\begin{tabular}{l l S[retain-explicit-plus] l}
		\toprule
		\multicolumn{2}{c}{Parameter}						&		\multicolumn{1}{c}{Value}	&	Direction\\
		\midrule
		&$g_{1}$	&	  2.00345+-0.00005	&	\hkl[110]\\
		&$g_{2}$	&		2.00274+-0.00005	&	\hkl[001]\\
		&$g_{3}$	&		2.00271+-0.00005 &	\hkl[1-10]\\
		\\
		\toprule
		\multicolumn{2}{c}{Parameter}	&		\multicolumn{1}{c}{Value (\si{\mega\hertz})}	&	Direction\\
		\midrule
		\Nfif&$A_{1}$	&	+3.47+-0.02			&	\SI{-3.5+-0.5}{\degree} from \hkl[11-2]\\
		&$A_{2}$	&		+4.51+-0.02			&	\SI{-3.5+-0.5}{\degree} from  \hkl[111]\\
		&$A_{3}$	&		+4.09+-0.02			& \hkl[1-10]\\	
		\\
		\Cthir&$A_{\parallel}$	&		+317.5+-0.5		&	\SI{2.0 +- 0.5}{\degree} from \hkl[-111]\\
		&$A_{\perp}$	&			+202.3+-0.5			&	\hkl[-1-10]\\
		\\		
		\Nfour&$P_{\parallel}$	&		-5.0+-0.1			&	\hkl[111]\\
		\bottomrule
	\end{tabular}
	\caption{The measured spin Hamiltonian parameters for the \NVNminus defect in \Nfif- and \Nfour-doped diamond. The sign of the nitrogen hyperfine interaction is determined by the negative quadrupolar interaction for \Nfour; the sign of the carbon hyperfine must be positive to account for the observed electron probability density localization. $g$-values referenced to \Nsneutral assuming isotropic $g=\num{2.0024}$. Directions relate to the defect orientation shown in figure~\ref{subfig:NVNdrawing} --- the hyperfine interaction for the second nitrogen / carbon can be generated by a \symspc{c_{2}} rotation about \hkl[001]. Angles are given as the acute angle between the interaction axis and \hkl[001].}
	\label{tab:spin_hamiltonian_parameters}
\end{table}

\section{Discussion}
\label{sec:discussion}

\subsection{The nitrogen hyperfine interaction}
\label{subsec:Nhyperfine}
In the simple case, the hyperfine interaction can be described as the sum of an isotropic component arising from non-zero spin density at the nucleus (Fermi contact), and an anisotropic dipole-dipole component. Typically the sign of the interaction is ambiguous: here, the sign of the \Nfour and thus \Nfif hyperfine was determined using knowledge of the sign of the quadrupole interaction (section~\ref{subsec:N14quadrupole}). 

The isotropic component $a = (A_{1} + A_{2} + A_{3})/3 = \SI[retain-explicit-plus]{+4.02}{\mega\hertz}$ (see table~\ref{tab:spin_hamiltonian_parameters}) is small and opposite in sign to that expected for a negative nuclear magnetic moment. This implies a negative unpaired electron probability density at the nitrogen nucleus, and hence indicates an indirect interaction (configuration interaction, exchange polarization):\cite{Watkins1975} a similar situation was found for the ground state of \NVminus, where the unpaired electron probability density is localized on three carbon neighbors.\cite{Felton2009a} For \NVNminus this suggests, as is confirmed by the \Cthir hyperfine interaction (section~\ref{subsec:electronwavefunction}), that the unpaired electron probability density is localized on the two carbon neighbors of \NVNminus, polarizing the core states of the nitrogen and, since the nuclear magneton for \Nfif is negative, yielding a positive Fermi contact term. Given the near-zero localization of the unpaired electron probability density on the nitrogen, the anisotropic dipole-dipole interaction must originate due to an interaction between the nitrogen and the unpaired electron probability density localized in an orbital on another atom.

A crude model was employed to aid interpretation of \Nfif anisotropic hyperfine interaction. In \NVNneutral, each nitrogen atom possesses two electrons in its lone pair pointing into the vacancy (each nitrogen is back-bonded to three carbons; see figure~\ref{subfig:NVN_structure}), whereas both carbon atoms neighboring the vacancy possess only one electron in the non-bonded orbital pointing towards the vacancy. These two orbitals interact to form an extended bonding orbital, which is thus fully occupied in \NVNneutral. It was therefore assumed that the unpaired electron probability density in \NVNminus is localized in an antibonding orbital formed between these two carbon atoms. An axial dipole-dipole interaction matrix of the form $b,b,-2b$ was constructed for each of the unpaired electron probability density locations, then each matrix was transformed into the crystal axes and summed to yield a macroscopic interaction matrix (the form measured in experiment). For each nitrogen atom the interaction was restricted to the \hkl{-1-10} plane containing the nitrogen, vacancy and the appropriate next-nearest neighbor carbon. No geometric relaxation was included in this model, and the unpaired electron probability density was taken to be localized at two points only.

 A least-squares fit was performed against the experimental values, with the only free parameters being $\theta$, the in-plane angle of the interaction and $b_{\mathrm{fit}}$, the strength of the interaction in MHz. $\theta$ was measured from the \hkl<0-1-1> direction between the nitrogen and the next-nearest neighbor carbon and gives the direction of the interaction; $b_{\mathrm{fit}}$ indicates the strength of the interaction (and hence dipole-dipole proximity). A value of $\theta = \SI{35.3}{\degree}$ would indicate that the unpaired electron probability density is localized on the axis connecting each nitrogen atom to the vacancy, whereas a value of $\theta = \SI{0}{\degree}$ suggests localization on the axis connecting each nitrogen atom to its next-nearest neighbor carbon atom.

\begin{table*}[htb]
	\centering
	\setlength{\tabcolsep}{8pt}
	\begin{tabular}{l l l}
		\toprule
				&		Magnitude (\si{\mega\hertz})	& Direction\\
		\midrule
		Calculation														&			$-0.58, +0.46, +0.12$								&			\hkl[11-2], \hkl[111], \hkl[1-10]  \\
		Experiment	&	$-0.55, +0.49, +0.07$	&	\SI{3.5}{\degree} from \hkl[11-2], \SI{3.5}{\degree} from \hkl[111], \hkl[110]\\
		\bottomrule
	\end{tabular}
	\caption{Results of calculations designed to minimize the difference between the calculated and observed hyperfine values. The values were calculated at an angle of $\theta=0$ (see text for details). Values can be compared to the experimental values given in table~\ref{tab:spin_hamiltonian_parameters} by adding the isotropic \SI{4.02}{\mega\hertz} component.}
	\label{tab:dipolar_hyperfine_calculations}
\end{table*}

The best fit to the dipolar interaction is given in table~\ref{tab:dipolar_hyperfine_calculations} and was achieved at an angle of $\theta=\SI{0}{\degree}$ and $b_{\mathrm{fit}}=\SI{-0.46}{\mega\hertz}$. The resulting macroscopic interaction differs in orientation from experiment by \SI{3}{\degree}. A simple electron-\Nfif dipolar calculation\cite{Schweiger2001} yields this value for $b_{\mathrm{fit}}$ with a \Nfif to carbon atom separation of approximately \SI{2.1}{\angstrom}. Reassuringly, this distance is only \SI{18}{\percent} smaller than the next-nearest neighbor separation of \SI{2.52}{\angstrom} in diamond. For a more precise calculation, information about shape and amplitude of the electron wavefunction is required in addition to knowledge about the minimum energy geometrical configuration.

\subsection{$\mathbf{^{13}C}$ hyperfine interaction and localization of the unpaired electron probability density}
\label{subsec:electronwavefunction}
The unpaired electron wavefunction $\Psi_{j}$ may written as a summation over all atoms where the unpaired electron probability density is non-zero:
\[
	\Psi_{j} = \sum_{n}^{N} \eta_{n} \psi_{n} \;\; \text{where} \;\; \sum_{n}^{N} \eta_{n}^{2} = 1 \;.
\] For carbon and nitrogen atoms in diamond, each atomic wavefunction $\psi_{n}$ consists of contributions from $s$- and $p$-type wavefunctions, with 
\[
	\psi_{n} = \alpha_{n} \phi_{2s} + \beta_{n} \phi_{2p} \;\; \text{and} \;\; \alpha_{n}^{2} + \beta_{n}^{2} = 1 \;.
\] Using standard tables and methods for hyperfine interpretation,\cite{Morton1978} the measured \Cthir hyperfine values $a=\SI{240.7}{\mega\hertz}$ and $b=\SI{38.4}{\mega\hertz}$ yield $\alpha_{\mathrm{C}}^{2}=\SI{15}{\percent}$, $\beta_{\mathrm{C}}^{2}=\SI{85}{\percent}$ and an unpaired electron probability density on each nearest-neighbor carbon atom of $\eta_{\mathrm{C}}^{2}=\SI{42}{\percent}$. 

The calculations above are consistent with one another and form a complete picture: the unpaired electron probability density is highly localized on the two carbon atoms, and the nitrogen interaction is weak due to virtually zero local unpaired electron probability density. Approximately \SI{85}{\percent} of the unpaired electron probability density can be accounted for by the two carbon atoms nearest-neighbor to the vacancy in \NVNminus. This is very similar to the \NVminus system, where \SI{84}{\percent} of the electron spin density can be accounted for by the three nearest-neighbor carbons,\cite{Felton2009a} with an $sp$-hybridization ratio of $\lambda^{2}=\num{6.2+-0.2}$ versus \num{5.6+-0.2} here; these values suggest that the relaxation of the carbons away from the vacancy is similar in both defects.

\subsection{The \texorpdfstring{$\mathbf{\Nfour}$}{14N} quadrupole interaction}
\label{subsec:N14quadrupole}
As discussed in the introduction, \NfourNVNminus{} has not previously been identified in \Nfour-doped diamond. However, its identification in \Nfif-doped diamond provides a route to identification in natural abundance samples: by scaling the hyperfine parameters by the ratio of the isotopic nuclear g-value, \cite{Stone2005} the only unknown parameter required to fit a potential \NfourNVNminus{} spectrum is the quadrupolar interaction with \Nfour. Using an electron localization in the \Nfour 2$p$ orbital of 0\%, as observed in experiment, a quadrupolar interaction strength of $P_{\parallel} \approx \SI{-5.2}{\mega\hertz}$ can be estimated following previous discussions on the magnitude of \Nfour quadrupolar interaction strengths in diamond.\cite{Tucker1994} Additionally, following other nitrogen- and vacancy-containing defects in diamond, the quadrupolar interaction can be assumed to be aligned along \hkl<111>.

\begin{figure}[btph]
	\centering
	\includegraphics[width=0.95\columnwidth]{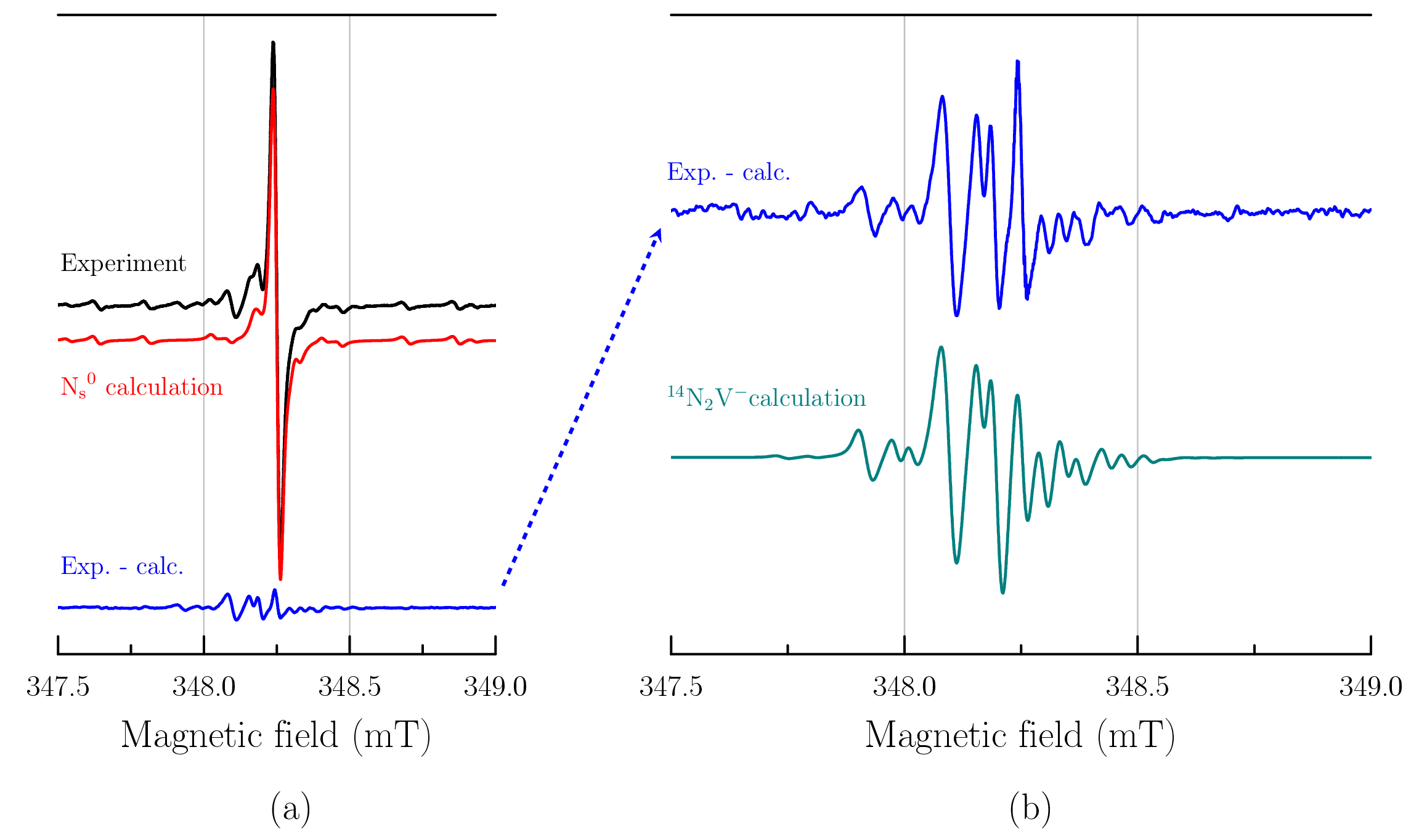}
	\caption{(a): experimental $B\|\hkl<001>$ EPR spectrum of a \Nfour-doped diamond post-irradiation and annealing; calculated spectrum of \NfourNSub and an additional line at $g = \num{2.00269}$; difference spectrum. (b): difference spectrum (as bottom left); \NfourNVNminus{} spectrum generated using the Hamiltonian parameters given in table~\ref{tab:dipolar_hyperfine_calculations}.}
	\label{fig:N14Comparisons}
\end{figure}

An HPHT synthetic sample with natural nitrogen isotope abundance was prepared using the treatment procedure described in \S\ref{sec:experiment}; subsequent measurements recorded the experimental spectrum shown in figure~\ref{fig:N14Comparisons}. Simultaneous fitting of \hkl<001> and \hkl<110> spectra yielded a best-fit value of $P_{\parallel} = \SI{-5.0}{\mega\hertz}$, in close agreement with the estimate based on other $\mathrm{N_{n}V}$ defects in diamond. The characteristic \Nfour hyperfine structure is not immediately apparent due to the large quadrupolar interaction leading to mixing of nuclear spin states (81 possible transitions per defect versus 16 per defect for \Nfif), with the effect that the spectrum is ``smeared out''. Additionally, the EPR spectra of \NfourNVNminus{} and \NfourNSub significantly overlap (and hence obscure one another --- see figure~\ref{fig:N14Comparisons}) but are spectrally separated in \Nfif-doped material (due to $I=1 \rightarrow I=\frac{1}{2}$ for \Nfour and \Nfif, respectively). A combination of the inherently more complex spectrum and the difficulty of producing high concentrations of \NVNminus{} without higher concentrations of \Nsneutral explains why \NfourNVNminus{} has not been previously identified.

\section{Conclusion}
\label{sec:conclusion}
An $S=\frac{1}{2}$ spectrum containing two near-\SI{100}{\percent} abundant $I=\frac{1}{2}$ nuclei has been observed in a \Nfif-doped synthetic diamond, indicating \Nfif as the nuclei involved. The corresponding defect has \RhombicI symmetry. A full description of the unpaired electron probability density localization has been proposed and is entirely consistent with all expected electronic and spectral attributes of \NVNminus, a rhombic defect in the $\mathrm{N_{n}V}$ family of diamond defects. The vast majority of unpaired electron probability density in the defect is distributed over the nearest-neighbor carbon atoms of the vacancy, as is the case in other vacancy-type defects in diamond.\cite{Tucker1994,Felton2009a}

The fitted spin Hamiltonian parameters were employed to identify the corresponding spectrum in an \Nfour-doped synthetic diamond, where the only free spectral parameter was the quadrupolar interaction strength --- the nuclei are therefore identified unambiguously with nitrogen. The large quadrupole interaction and overlap of the \NfourNVNminus{} and \NfourNSub EPR spectra explains why the defect has not previously been identified.

The integrated intensity of the \NVNminus EPR spectrum has been shown to correlate with the integrated absorption of the H2 optical band over an order of magnitude in concentration, further strengthening the assignment of H2 to the \NVNminus defect.

Isotopic enrichment has proved instrumental in the understanding of \NVNnb, and demonstrates the importance of complementary synthesis techniques in developing a holistic view of the possibilities of defect engineering in diamond. Isotopic enrichment may be key in the exploitation of this defect as a photochromic memory with EPR readout.\cite{Mita1990}

%

\end{document}